\documentclass[reprint,amsmath,amssymb,aps]{revtex4-1}

\usepackage{graphicx}
\usepackage{dcolumn}
\usepackage{bm}

\usepackage{booktabs}
\usepackage{mhchem}
\usepackage{url}

\begin{document}

\title{Quantum deep field: data-driven wave function, electron density generation,
and atomization energy prediction and extrapolation with machine learning}

\author{Masashi Tsubaki}
\altaffiliation{National Institute of Advanced Industrial Science and Technology}
\email{tsubaki.masashi@aist.go.jp}

\author{Teruyasu Mizoguchi}
\altaffiliation{Institute of Industrial Science, the University of Tokyo}
\email{teru@iis.u-tokyo.ac.jp}

\begin{abstract}
Deep neural networks (DNNs) have been used to successfully predict molecular properties
calculated based on the Kohn--Sham density functional theory (KS-DFT).
Although this prediction is fast and accurate, we believe that
a DNN model for KS-DFT must not only predict the properties
but also provide the electron density of a molecule.
This letter presents the quantum deep field (QDF),
which provides the electron density with an unsupervised
but end-to-end physics-informed modeling
by learning the atomization energy on a large-scale dataset.
QDF performed well at atomization energy prediction,
generated valid electron density, and demonstrated extrapolation.
Our QDF implementation is available at
\url{https://github.com/masashitsubaki/QuantumDeepField_molecule}.
\end{abstract}

\pacs{Valid PACS appear here}

\maketitle

Quantum chemical simulations, such as
Kohn--Sham density functional theory (KS-DFT) calculations
have been recently approximated by machine learning (ML) techniques
such as kernel methods~\cite{rupp2012fast,brockherde2017bypassing,faber2017prediction}.
Very recently, deep neural networks (DNNs)~\cite{schutt2017quantum,schutt2018schnet}
have been used to successfully predict molecular properties,
such as the atomization energy, HOMO, and LUMO, on large-scale datasets.
Although this prediction is fast and accurate, there is a problem:
a DNN model for KS-DFT must be consistently based on an understanding of
the underlying physics and must not only predict the properties
but also provide the fundamental quantum characteristics
(i.e., the wave function/orbital and electron density) of a molecule.
Most existing DNN models, however, cannot provide the electron density because
they consider only the atomic coordinates and ignore the molecular field.
Furthermore, they are mainly interested in predicting the final output,
i.e., the interpolation accuracy of a molecular property within a benchmark dataset,
and do not focus on capturing the fundamental characteristics of molecules.
This does not lead to learning a physically meaningful model
and extrapolation, i.e., prediction for totally unknown molecules in terms of
its size and structure that do not appear in the dataset.
Extrapolation is important in not only molecular science
but also real applications for transferring to other molecules or crystals
and predicting their properties~\cite{huan2016polymer,jain2013commentary}
in materials informatics.

\begin{figure}[t]
\begin{center}
\includegraphics[width=8.5cm, bb=0 0 1024 440]{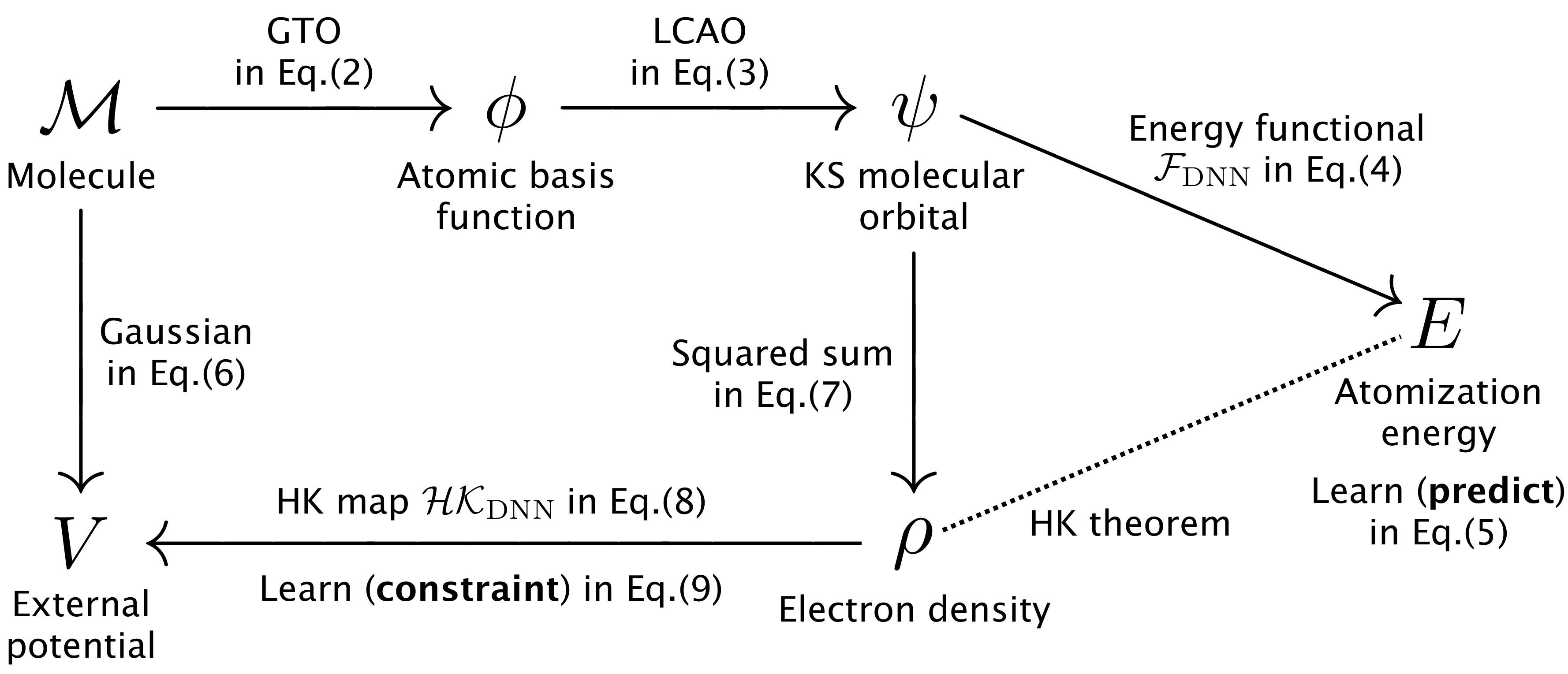}
\caption
{\label{overview}
An overview of the computational flow of our proposed QDF framework.
The details of each arrow in this figure are described in the corresponding equation.
}
\end{center}
\end{figure}

\begin{figure*}[t]
\begin{center}
\includegraphics[width=17cm, bb=0 0 1920 596]{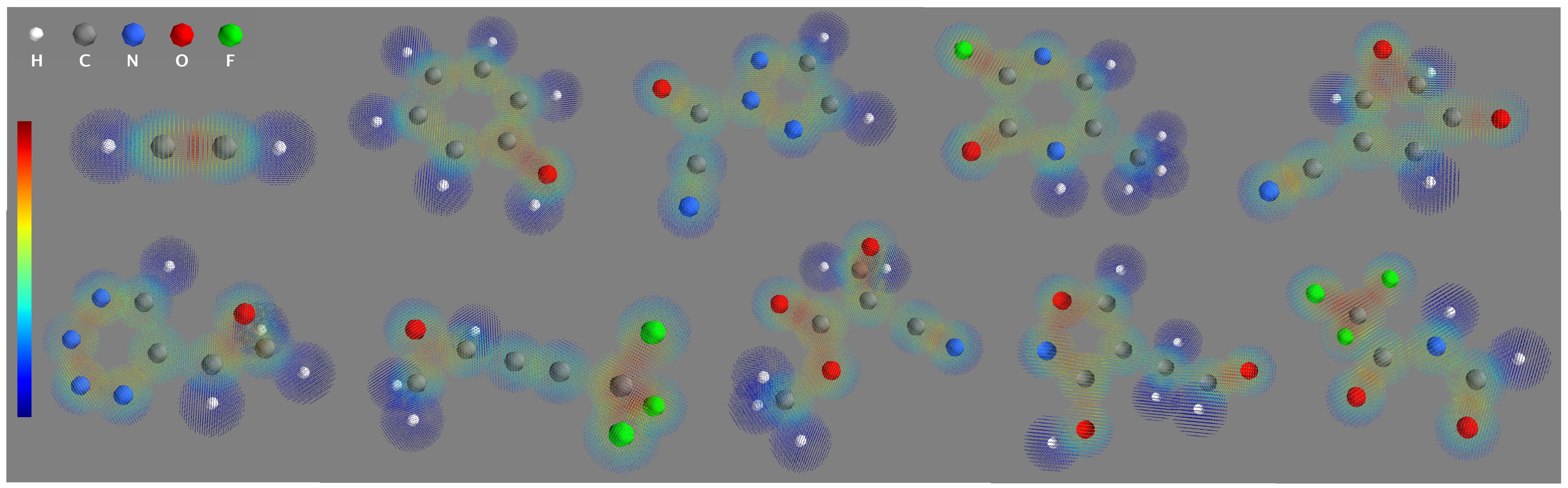}
\caption
{\label{densitymap}
The electron density maps of some molecules generated by QDF.
Each map can be obtained within a second.
}
\end{center}
\end{figure*}

In this letter, we present a simple framework called the quantum deep field (QDF)
that provides the electron density of molecules
by learning their atomization energies on a large-scale dataset.
Crucially, our data-driven QDF framework requires only
the molecule--energy pairs (e.g., the QM9 dataset~\cite{ramakrishnan2014quantum})
and does not require the molecule--density pairs for training;
in other words, QDF generates the electron density indirectly or
in an unsupervised fashion with end-to-end physics-informed modeling.
The QDF model involves three linear/nonlinear components:
(1)~a linear combination of atomic orbitals (LCAO):
$\phi \rightarrow \psi$, where $\phi$ is the atomic basis function given by
the Gaussian-type orbital (GTO) and $\psi$ is the KS molecular orbital,
(2)~a nonlinear energy functional: $\psi \rightarrow E$, where $E$ is the atomization energy,
and (3)~a nonlinear Hohenberg--Kohn (HK) map: $\rho \rightarrow V$,
where $\rho$ is the electron density and $V$ is the external potential.
We optimize all parameters of the LCAO, energy functional, and HK map,
in which the latter two are implemented by simple DNNs, simultaneously
using the backpropagation and stochastic gradient descent (SGD)~\cite{kingma2014adam}.
We emphasize that learning $\psi \rightarrow \rho \rightarrow V$ serves as
the physical constraint on learning $\psi \rightarrow E$ (FIG.~\ref{overview});
this allows QDF to provide valid $\psi$ and $\rho$ (FIG.~\ref{densitymap})
and leads to high prediction and better extrapolation performance on $E$.

Initially, a molecule is denoted by
$\mathcal M = \{ (a_1, \mathbf R_1), (a_2, \mathbf R_2),
\cdots, (a_M, \mathbf R_M) \} = \{ (a_m, \mathbf R_m) \}_{m=1}^M$,
where $a_m$ is the $m$th atom,
$\mathbf R_m$ is the 3D coordinate of $a_m$,
and $M$ is the number of atoms in $\mathcal M$.
Given $\mathcal M$, a set of the molecular orbitals (or wave functions)
is denoted by $\{ \psi_1(\mathbf r), \psi_2(\mathbf r),
\cdots, \psi_N(\mathbf r) \} = \{ \psi_n(\mathbf r) \}_{n=1}^N$,
where $\mathbf r$ is a position in the field
and $N$ is the number of orbitals.
The LCAO (or the superposition of wave functions)
provides the $n$th molecular orbital:
\begin{eqnarray}
\label{MO}
\psi_n(\mathbf r) = \sum_{i=1}^N c_{ni} \phi_i(\mathbf r - \mathbf R_i)
\:\:\: \text{s.t.} \:\:\: \sum_{i=1}^N c_{ni}^2 = 1,
\end{eqnarray}
where $c_{ni}$ is the $i$th coefficient,
$\phi_i(\mathbf r - \mathbf R_i)$ is the $i$th atomic basis function
whose origin is $\mathbf R_i$, and $N$ is the number of basis functions.
As the basis function, we use the GTO:
\begin{eqnarray}
\label{GTO}
\phi_i(\mathbf r - \mathbf R_i)
= \frac{1}{Z(q_i, \zeta_i)} D_i^{(q_i-1)} e^{-\zeta_i D_i^2},
\end{eqnarray}
where $D_i = || \mathbf r - \mathbf R_i ||$,
$q_i$ is the principle quantum number, $\zeta_i$ is the orbital exponent,
and $Z(q_i, \zeta_i)$ is the normalization term.
Additionally, we use the 6-31G basis set and represent
$\{ \psi_n(\mathbf r) \}_{n=1}^N = \boldsymbol \psi(\mathbf r)$
with the $N$-dimensional vector:
\begin{eqnarray}
\label{LCAO}
\boldsymbol \psi(\mathbf r) =
\sum_{i=1}^N \mathbf c_i \phi_i(\mathbf r - \mathbf R_i),
\end{eqnarray}
where $\boldsymbol \psi(\mathbf r) \in \mathbb R^N$
has $\psi_n(\mathbf r)$ as its $n$th element
and $\mathbf c_i \in \mathbb R^N$ has $c_{ni}$ as its $n$th element.
Note that the orbital exponents $\{ \zeta_1, \zeta_2, \cdots,
\zeta_N \} = \{ \zeta_i \}_{i=1}^N$
and the coefficient vectors $\{ \mathbf c_1, \mathbf c_2, \cdots,
\mathbf c_N \} = \{ \mathbf c_i \}_{i=1}^N$ are randomly initialized
and then learned/optimized for predicting the atomization energy
using the backpropagation and SGD.

For $\boldsymbol \psi$, we express a DNN-based
energy functional $\mathcal{F}_{\text{DNN}}$ as follows:
\begin{eqnarray}
\label{FDNN}
E'_{\mathcal M} = \mathcal F_\text{DNN}[\boldsymbol \psi],
\end{eqnarray}
where $E'_{\mathcal M}$ is the predicted atomization energy of $\mathcal M$.
Herein we use a simple feedforward architecture
for the implementation of $\mathcal F_\text{DNN}$, in which
the DNN models the interaction between $\psi_n$ and $\psi_m$.
Finally, we minimize the loss function:
\begin{eqnarray}
\label{L_E}
\mathcal L_E = || E_{\mathcal M} - E'_{\mathcal M}||^2,
\end{eqnarray}
where $E_{\mathcal M}$ is the atomization energy of $\mathcal M$ in the QM9 dataset.
Details about $\mathcal{F}_{\text{DNN}}$ and its optimization
are described in Supplementary.

Unfortunately, only minimizing $\mathcal L_E$ does not lead to
learning a physically meaningful model; because the DNN has a strong nonlinearity,
such a powerful $\mathcal F_\text{DNN} [\boldsymbol \psi]$
will output the correct atomization energy $E_{\mathcal M}$
even if $\boldsymbol \psi$ is not valid.
In other words, the model does not guarantee the KS orbitals
$\boldsymbol \psi(\mathbf r)$, which means that
$\boldsymbol \psi(\mathbf r)$ cannot provide the correct electron density by
$\rho(\mathbf r) = \sum_{n=1}^N |\psi_n(\mathbf r)|^2$.
To address this problem, we impose a constraint on
$\boldsymbol \psi(\mathbf r)$ based on the HK theorem,
which ensures that the external potential $V(\mathbf r)$ is a unique
(i.e., a nonlinear but one-to-one correspondence) function
of the electron density $\rho(\mathbf r)$,
i.e., $V(\mathbf r) \leftrightarrow \rho(\mathbf r)$.
Specifically, we implement the constraint by a nonlinear map
$\rho(\mathbf r) \rightarrow V(\mathbf r)$,
which we refer to as the HK map~\cite{brockherde2017bypassing,moreno2020deep}
and learn the nonlinearity using a simple DNN.

Formally, we consider a Gaussian external
potential~\cite{bartok2010gaussian,brockherde2017bypassing}:
\begin{eqnarray}
\label{V}
V_{\mathcal M}(\mathbf r) =
- \sum_{i=m}^M Z_m e^{- ||\mathbf r - \mathbf R_m ||^2},
\end{eqnarray}
where $Z_m$ is the nuclear charge of $a_m$.
We assume $V_{\mathcal M}(\mathbf r)$ to be the correct
external potential of $\mathcal M$; that is, $V_{\mathcal M}(\mathbf r)$
is used as a target for minimizing loss in the model.
Additionally, the electron density is given by
\begin{eqnarray}
\label{rho}
\rho(\mathbf r) = \sum_{n=1}^N | \psi_n(\mathbf r) |^2.
\end{eqnarray}
For $\rho(\mathbf r)$, we express a DNN-based
HK map $\mathcal{HK}_{\text{DNN}}$ as follows:
\begin{eqnarray}
\label{HKDNN}
V'_{\mathcal M}(\mathbf r) = \mathcal{HK}_\text{DNN} (\rho(\mathbf r)),
\end{eqnarray}
where $V'_{\mathcal M}(\mathbf r)$ is the predicted external potential of $\mathcal M$.
This letter uses a simple feedforward architecture
for the implementation of $\mathcal{HK}_\text{DNN}$.
Finally, we minimize the loss function:
\begin{eqnarray}
\label{L_V}
\mathcal L_V = ||V_{\mathcal M}(\mathbf r) - V'_{\mathcal M}(\mathbf r)||^2.
\end{eqnarray}
Details about $\mathcal{HK}_\text{DNN}$ and its optimization
are described in Supplementary.

Note that the HK map used in~\cite{brockherde2017bypassing}
learns $V(\mathbf r) \rightarrow \rho(\mathbf r)$
by a supervised kernel method, where $V(\mathbf r)$ is the input potential
and $\rho(\mathbf r)$ is the target density to be learned.
In contrast, our HK map is different; that is, the direction is opposite,
i.e., $\rho(\mathbf r) \rightarrow V(\mathbf r)$,
where the input density is $\rho(\mathbf r) = \sum_{n=1}^N |\psi_n(\mathbf r)|^2$,
$\psi_n(\mathbf r)$ is obtained by LCAO,
and $V(\mathbf r)$ is the target potential to be learned.

Furthermore, please note that as a total learning algorithm of QDF, we minimize
Eq.~(\ref{L_E}) and Eq.~(\ref{L_V}) alternately in an end-to-end fashion
and optimize all parameters of the LCAO,
$\mathcal{F}_{\text{DNN}}$, and $\mathcal{HK}_{\text{DNN}}$
using the backpropagation and SGD.
We believe that this algorithm involving the HK map constraint
allows QDF to function as a self-consistent learning machine for KS-DFT.

\begin{figure}[t]
\begin{center}
\includegraphics[width=8.5cm, bb=0 0 1920 744]{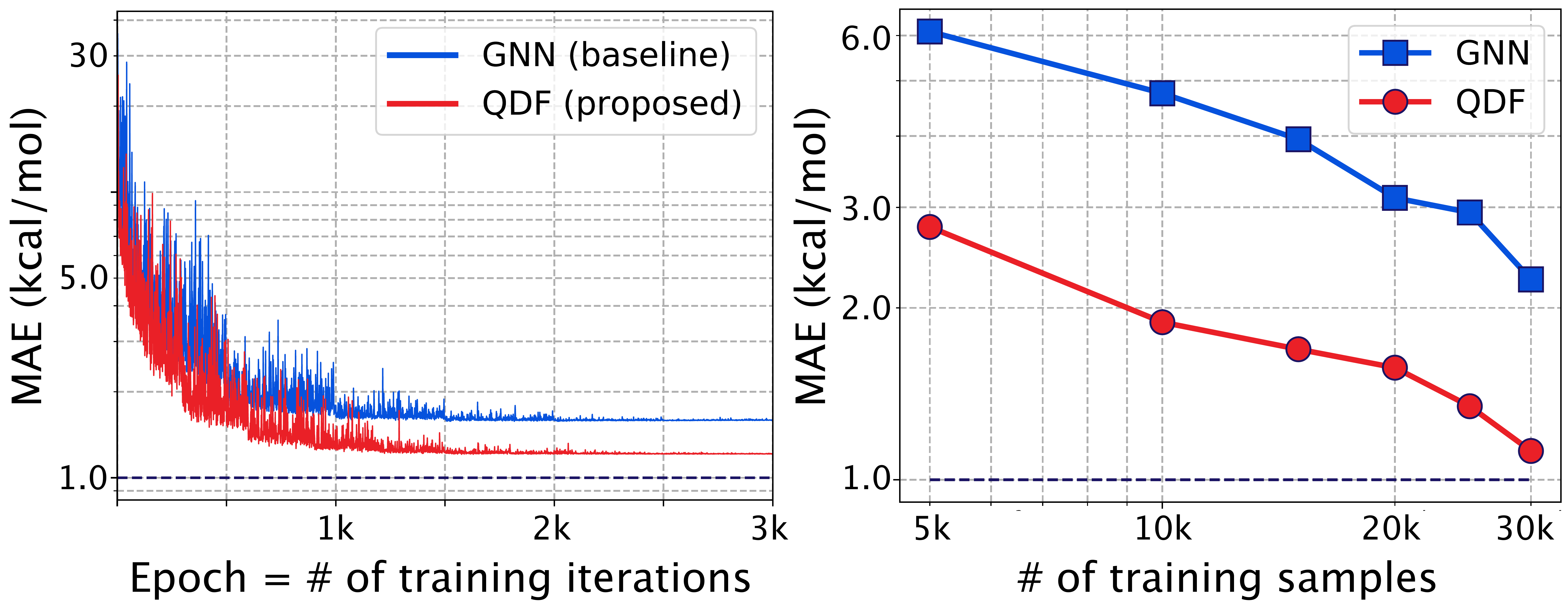}
\caption
{\label{learningcurve}
(a)~The learning curves of GNN and QDF for atomization energy prediction,
which were performed on the QM9under14atoms dataset.
(b)~The number of training molecules versus MAE.
The number of test samples was fixed at 10,000 molecules
and we varied the number of training molecules,
which were extracted from the QM9 dataset.
}
\end{center}
\end{figure}

\begin{table}[t]
\begin{center}
\begin{tabular}{lrr}
\hline \hline
Model & \# of parameters & MAE (kcal/mol) \\
\hline
GNN (baseline) & 483,631 & 1.58 \\
DTNN~\cite{schutt2017quantum} & --- & 1.51 \\
SchNet~\cite{schutt2018schnet} & 1,676,133 & 1.23 \\
QDF (proposed) & 495,262 & 1.21 \\
\hline
Chemical accuracy & & 1.00 \\
\hline \hline
\end{tabular}
\end{center}
\caption{
\label{modelsize}
The model sizes and final prediction errors on the QM9under14atoms dataset.
The number of parameters and MAE for SchNet were obtained
from the SchNetPack of original paper~\cite{schutt2018schnet}.
We note that these results can vary and SchNet may outperform QDF
with careful tuning of its hyperparameters; however,
our main aim herein is not to build a competitive model
with regard to the interpolation performance within a single benchmark dataset.
}
\end{table}

\begin{figure}[t]
\begin{center}
\includegraphics[width=8.5cm, bb=0 0 1920 1080]{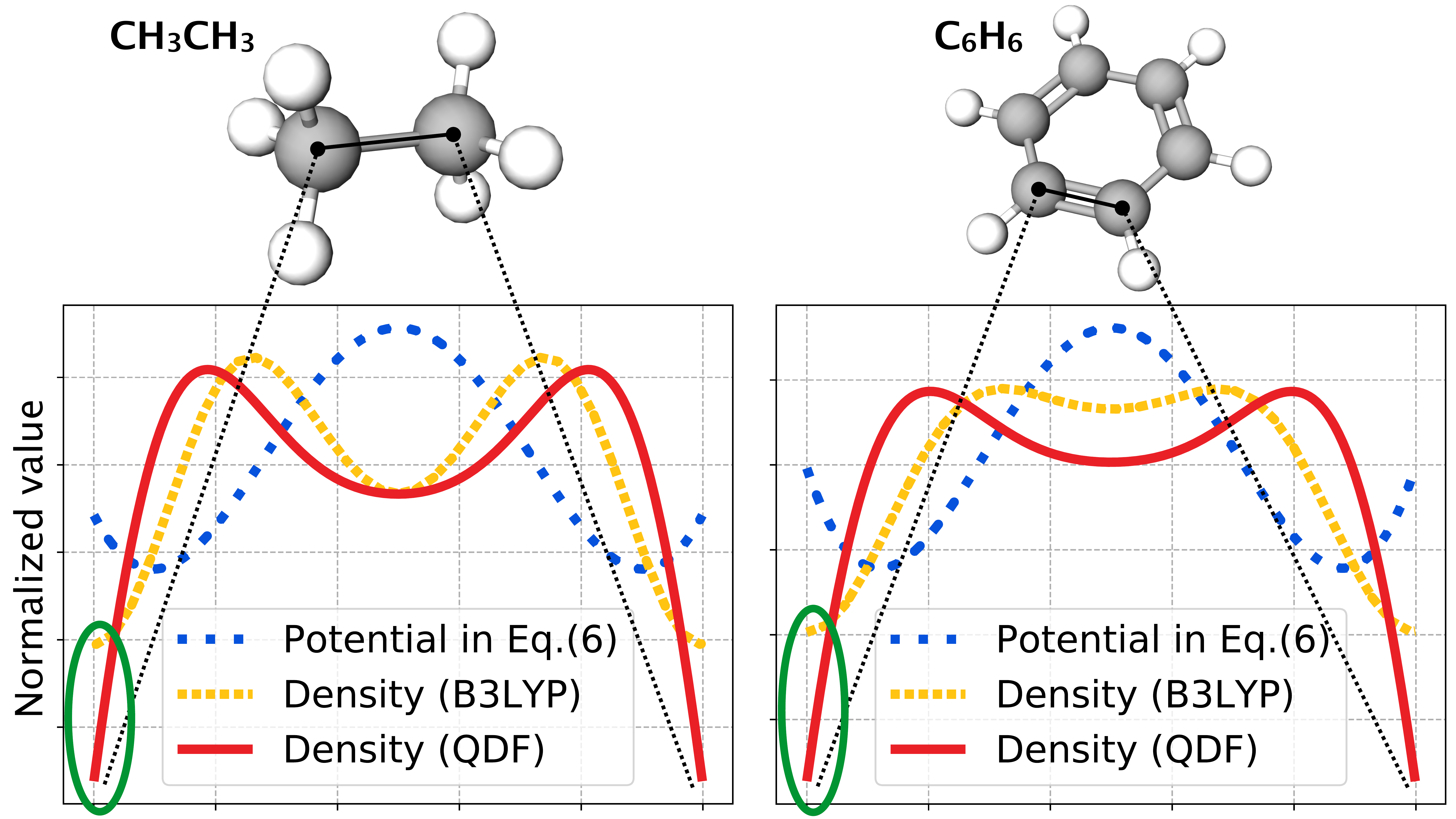}
\caption
{\label{densitypeak}
The electron densities, generated by our QDF
and calculated by the B3LYP simulation, on the chemical bonds.
We extracted 1,000 points between the two atoms on the x-axis
and displayed the intensity (i.e., the normalized density or potential value) on the y-axis.
}
\end{center}
\end{figure}

\begin{figure*}[t]
\begin{center}
\includegraphics[width=17cm, bb=0 0 1920 658]{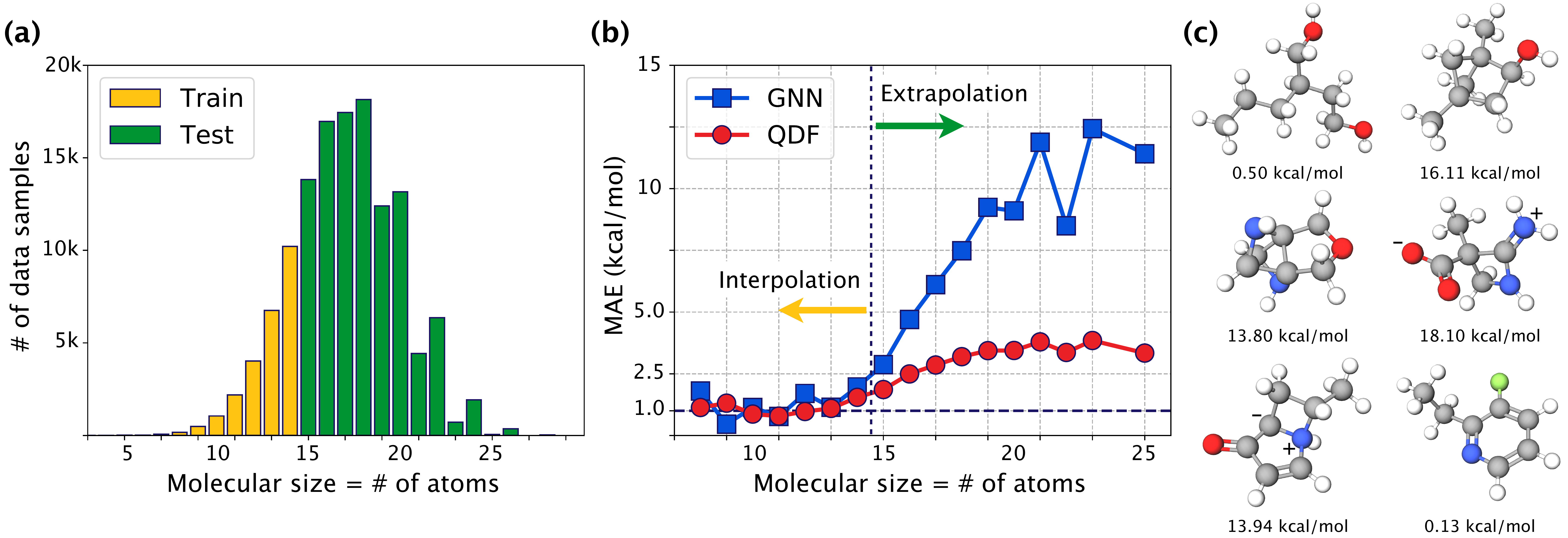}
\caption
{\label{extrapolation}
(a)~The data distribution of interpolation (train) and extrapolation (test) samples.
The number of test molecules of the QM9over15atoms dataset is 115,000,
which is 10-times more than that of training molecules of the QM9under14atoms dataset.
(b)~The MAEs on this large-scale extrapolation evaluation.
The extrapolation error (the right side of this figure)
of our QDF is approximately 3.0 kcal/mol.
(c)~Examples of predicted molecules by QDF and their errors.
}
\end{center}
\end{figure*}

To evaluate the QDF performance, we first describe
the learning/prediction results of atomization energy $E$.
For training and testing, we used the QM9under14atoms dataset,
which is a subset (15,000 samples of relatively small-sized molecules)
of the full set (130,000 samples) of the QM9 dataset (see Supplementary).
FIG.~\ref{learningcurve}(a) displays the prediction errors,
mean absolute error (MAE, where lower is better) in units of kcal/mol,
of our QDF and a baseline model in the form of a learning curve.
As the baseline, we implemented a variant of graph neural networks
(GNNs)~\cite{kearnes2016molecular}, which uses the molecular structure
(i.e., the types and 3D coordinates of constituent atoms) alone,
does not consider the molecular field,
and has hierarchical structure and strong nonlinearity
in the modeling phase of the molecular structure.
Compared to the GNN, our QDF was able to predict $E$
and its error was much closer to chemical accuracy (1.0 kcal/mol).
Additionally, FIG.~\ref{learningcurve}(b) describes
how the accuracy improves as the number of training samples increases.
The resulting curve was almost linear, and this result is useful for
estimating how many training samples are required to achieve a desired accuracy.

TABLE~\ref{modelsize} shows the final prediction errors and model sizes of
the GNN, QDF, and others~\cite{schutt2017quantum,schutt2018schnet} as references.
SchNet, which is a variant of the earlier proposed deep tensor neural network (DTNN),
is now a standard state-of-the-art deep learning model.
We argue that our implemented GNN is not a weak baseline because
it achieved a reasonable performance that was competitive with that of DTNN.
Additionally, we believe that QDF outperformed (or was competitive with)
SchNet in terms of the prediction error; both MAEs were close to 1.20 kcal/mol.
In terms of the model size, however, SchNet has more than 1.5 million learning parameters;
in contrast, QDF, which has less than half a million parameters, is much more compact.

From the test dataset, we extracted 10 molecules
and visualized their electron density maps in FIG.~\ref{densitymap}
using Mayavi~\cite{ramachandran2011mayavi}.
We first observed that the electron density $\rho$ is higher at points around
C, N, O, and F, which have high electronegativity (red to green)
than that at points around H, which has low electronegativity (blue),
indicating that $\rho$ transfers from H to C, N, O, and F
were qualitatively reproduced by learning the atomization energy $E$.
Additionally, $\rho$ increases between two atoms; for example,
$\rho$ between two carbon atoms in \ce{C2H2}, \ce{C6H5OH},
and other molecules are high, indicating that
the formation of double and triple bonding was also reproduced.
Furthermore, $\rho$ is much higher on the bonding of F atoms.
We believe that QDF has the potential to generate valid $\rho$
even if the model is trained only with respect to $E$.

We compared the electron densities obtained by
our QDF and a hybrid-functional (B3LYP) simulation, in which
we focused on the density profiles along the chemical bonds
of ethane (\ce{CH3CH3}) and benzene (\ce{C6H6}) shown in FIG.~\ref{densitypeak}.
We first found that the QDF density (red line) could capture the characteristic
double peak features of the B3LYP density (yellow dotted line)
on the \ce{C-C} bond in \ce{CH3CH3} and the \ce{C=C} bond in \ce{C6H6}.
Additionally, comparing \ce{CH3CH3} and \ce{C6H6}, the difference
with respect to the sharpness of these double peaks could also be reproduced.

Furthermore, we compared the net charge of an atom in a molecule
obtained by our QDF with that obtained by the B3LYP simulation
and estimated by Bader analysis~\cite{tang2009grid}.
Indeed, Bader analysis cannot be directly applied to the current QDF
because we generated the grid points around the atoms assuming
a spherical distribution (see FIG.~\ref{densitymap} and Supplementary).
Therefore, we estimated the net charge of the \ce{C} atom
by summing the electron densities inside a sphere with a specific radius,
which is half the length of the \ce{C-C} bond in \ce{CH3CH3} (0.76 \AA)
and the \ce{C=C} bond in \ce{C6H6} (0.70 \AA).
Each estimated net charge is as follows:
\ce{C} in \ce{CH3CH3}: 4.05 (B3LYP by Bader), 3.87 (QDF by sphere), and its error is $-4.44\%$;
\ce{C} in \ce{C6H6}: 4.11 (B3LYP by Bader), 3.89 (QDF by sphere), and its error is $-5.35\%$.
These indicate that our QDF could quantitatively reproduce
the electron density distribution when compared with that calculated by the B3LYP.

Here, note that we simply followed the work of
Brockherde et al.~\cite{brockherde2017bypassing}
and used the Gaussian external potential as a target for learning the HK map.
Although the external potential $V$ is not limited to a Gaussian,
the current $V$ between two atoms (blue dotted line in FIG.~\ref{densitypeak})
seems to be reasonable for reproducing the double peak of the B3LYP density.
We also believe that this Gaussian potential should be improved
because our results could not reproduce the density
close to the nucleus (green circle in FIG.~\ref{densitypeak}).
Hence, another $V$ and atomic orbital,
e.g., a Slater-type orbital (STO), would be more suitable.
Overall, FIG.~\ref{densitypeak} demonstrates
the viability of QDF for generating valid $\rho$ in an unsupervised fashion.

Lastly, using a more practical evaluation setting,
we present evidence that QDF can capture
the molecular orbital/wave function and electron density,
i.e., the fundamental characteristics of molecules, from a large-scale dataset.
We believe that the following result is an interesting finding
of this study that has the potential to facilitate extrapolation,
which is difficult to solve by general ML approaches in principle.

We assume that if an ML model could capture the fundamental characteristics of data,
the model can be used to conduct a prediction
for totally unknown data, i.e., perform an extrapolation.
This study evaluated an extrapolation as follows:
we trained a model with small molecules and then tested it with large molecules.
Specifically, we trained the QDF model with small molecules
consisting of fewer than 14 atoms (15,000 samples; the prediction performance
was already shown in FIG.~\ref{learningcurve}(a) and TABLE~\ref{modelsize})
and then tested it with large molecules consisting of more than 15 atoms
(115,000 samples), which is the remainder of the QM9 dataset
(see FIG.~\ref{extrapolation}(a)).

FIG.~\ref{extrapolation}(b) shows the results of this large-scale extrapolation evaluation.
The accuracy achieved by GNN is the same as (or superior to)
that of the QDF in interpolation; however, the QDF could maintain this accuracy
even when the molecular size increases in extrapolation, whereas the GNN could not.
Actually, the GNN and its variants have strong nonlinearity
in the modeling phase of the molecular structure.
By eliminating such nonlinearity using LCAO and imposing the HK map constraint,
QDF does not suffer from overfitting.
We believe that this is evidence that QDF can capture
the fundamental quantum characteristics of molecules that are independent of the system.

FIG.~\ref{extrapolation}(c) shows the error analysis on QDF.
We found that the molecules with large errors (over 10 kcal/mol)
are often sterically strained or have a cage-like structure and are polarized.
However, the number of such molecules in the QM9 dataset is very small,
e.g., the ratio of polarized molecules is less than 0.3\%.
In contrast, although the ratio of molecules including F atom(s) is less than 1.5\%,
their errors were not large, approximately 0.1 kcal/mol,
because these molecules have a stable ring structure and are not polarized;
this shows the robustness of QDF.

This study proposed the QDF framework, which is different from other
deep learning approaches; in other words, QDF is not an extension of existing DNN models.
Our aim was to design a simple DNN model without incorporating
complicated techniques/architectures.
To model the electron density, some ML approaches have been
proposed~\cite{grisafi2018transferable,zhang2019embedded,
gong2019predicting,chandrasekaran2019solving,ryczko2019deep};
in particular, we were inspired by Grisafi et al.~\cite{grisafi2018transferable}.
They (1)~created an electron density dataset including
1,000 configurations of a few kinds of small molecules (\ce{C4H6} and \ce{C4H10}),
(2)~learned a Gaussian process (GP) model for $\rho$
(i.e., supervised learning) of these molecules,
and (3)~evaluated the transferability of the learned model
by predicting the $\rho$ of large molecules (\ce{C8H10} and \ce{C8H18}).
In contrast, our QDF can be viewed as an unsupervised model to reproduce $\rho$
using only a large-scale dataset of atomization energy $E$ (not $\rho$).
Additionally, we demonstrated the large-scale extrapolation in predicting $E$.
A supervised model as GP would be relatively easy to train
and superior to QDF in predicting $\rho$ because the unsupervised QDF is a model
closer to generative adversarial networks~\cite{goodfellow2014generative}
and would be unstable when generating $\rho$.

Furthermore, QDF can also be viewed as one of the approaches
such as the physics-informed, Hamiltonian, Fermionic neural
networks~\cite{raissi2019physics,pun2019physically,greydanus2019hamiltonian,pfau2019ab};
these solve the physical problems and equations using physically meaningful modeling.
QDF is designed as a self-consistent machine to solve
the KS equation with minimal (three) learning and physical constraint components:
LCAO, $\mathcal F_\text{DNN}[\boldsymbol \psi]$, and $\mathcal{HK}_\text{DNN}(\rho)$.
We believe that integrating a supervised model with a dataset of the electron
density~\cite{sinitskiy2018deep} (i.e., $\rho$ in FIG.~\ref{overview} is given as a target)
and an unsupervised but physically informed and meaningful model with a dataset
of the atomization energy, HOMO--LUMO gap, and other properties~\cite{chen2019alchemy}
will yield an interesting hybrid ML model.
QDF will admit many extensions (e.g., for crystals~\cite{xie2018crystal})
and applications (e.g., for transfer learning to solve more practical problems
in materials informatics~\cite{huan2016polymer,lopez2016harvard,kim2017hybrid}),
and our research position and future directions could prove useful (see Supplementary).

\bibliography{references}

\section*{Supplementary}

\subsection*{Dataset}

The QDF model was trained using the QM9 dataset~\cite{ramakrishnan2014quantum},
which contains approximately 130,000 samples of small, stable organic molecules
made up of H, C, N, O, and F atoms, along with 13 quantum chemical properties
(e.g., the atomization energy, HOMO, and LUMO) for each molecule.
These molecular properties were obtained by a DFT calculation
(Gaussian 09) at the B3LYP/6-31G(2df,p) level of theory.
However, the model could not be trained using all 130,000 molecules
owing to the computational cost of processing
a large number of grid points in each molecular field.
Therefore, this study used a subset of the QM9 dataset
with a limited number of atoms $M \leqq 14$ per molecule,
which we refer to as the QM9under14atoms dataset in the main text.
The number of samples in the QM9under14atoms dataset was approximately 15,000 molecules.
We randomly shuffled and split the QM9under14atoms into
training/validation/test sets as 8/1/1, in which
the validation set was used to tune the model hyperparameters
(see the Hyperparameters section below).
We calculated the MAE in TABLE~I (in the main text) by taking the mean of all results
obtained by 10-fold cross validation (CV) on the QM9under14atoms dataset.
Additionally, we chose a result from the CV results
and display its learning curve in FIG.~3(a) in the main text.
For the extrapolation evaluation, we used the QM9over15atoms dataset
(i.e., $M \geqq 15$), in which the number of samples was approximately 115,000 molecules,
which is 10 times more than that of the QM9under14atoms (training) dataset.

We emphasize that since QDF requires only the molecule--energy
(or other properties) pairs for training, we can use the popular QM9 dataset.
This has the advantage that existing large-scale datasets can be easily leveraged
and we could use another one such as the Alchemy dataset~\cite{chen2019alchemy}.
Very recently, a large-scale electron density dataset was
created~\cite{sinitskiy2018deep} for supervised learning; however,
we believe that QDF, which is an unsupervised method for learning electron density,
is a reasonable proposition, and integrating the supervised/unsupervised approaches
with the density/energy datasets could lead to an interesting hybrid model in the future.

\subsection*{Molecular field definition}

Given a molecule $\mathcal M = \{ (a_m, \mathbf R_m) \}_{m=1}^M$,
we consider spheres with a radius of $S$~\AA,
where each sphere covers each atom centred on $\mathbf R_m$, and then
divide the sphere into grids (or meshes) in intervals of $G$~\AA.
This process yields many grid points in $\mathcal M$,
as shown in FIG.~2 in the main text.
The grid-based field of $\mathcal M$ is denoted by
$\mathcal F_{\mathcal M} = \{ \mathbf r_1, \mathbf r_2, \cdots,
\mathbf r_{F_{\mathcal M}} \} = \{ \mathbf r_j \}_{j=1}^{F_{\mathcal M}}$,
where $\mathbf r_j$ is the 3D coordinate of the $j$th point
and $F_{\mathcal M}$ is the number of points in $\mathcal F_{\mathcal M}$.

\subsection*{Architectures of $\mathcal F_\text{DNN}$ and $\mathcal{HK}_\text{DNN}$}

To implement $\mathcal F_\text{DNN}$ described in the main text,
given $\{ \boldsymbol \psi(\mathbf r_j) \}_{j=1}^{F_{\mathcal M}}$,
where each $\boldsymbol \psi(\mathbf r_j)$ denotes the KS orbitals
obtained by LCAO on $\mathbf r_j$ in the above defined grid field,
we first consider the following feedforward architecture:
\begin{eqnarray}
\boldsymbol \psi^{(\ell+1)}(\mathbf r_j) = \text{ReLU} ( \mathbf W_E^{(\ell)}
\boldsymbol \psi^{(\ell)}(\mathbf r_j) + \mathbf b_E^{(\ell)} ),
\end{eqnarray}
where $\ell = 1, 2, \cdots, L$ is the number of hidden layers
($\boldsymbol \psi^{(1)}(\mathbf r_j) = \boldsymbol \psi(\mathbf r_j)$
and $L$ is the final layer), ReLU is the nonlinear activation function
$\text{ReLU}(\mathbf x) = \max(0, \mathbf x)$,
$\mathbf W_E^{(\ell)} \in \mathbb R^{N \times N}$ is the weight matrix in layer $\ell$,
and $\mathbf b_E^{(\ell)} \in \mathbb R^N$ is the bias vector in layer $\ell$.
We then sum over $\{ \boldsymbol \psi^{(L)}(\mathbf r_j) \}_{j=1}^{F_{\mathcal M}}$
and output an atomization energy with the following vanilla linear regressor:
\begin{eqnarray}
E'_{\mathcal M} = \mathbf w_E^{\top} \Big( \sum_{j=1}^{F_{\mathcal M}}
\boldsymbol \psi^{(L)}(\mathbf r_j) \Big) + b_E,
\end{eqnarray}
where $\mathbf w_E \in \mathbb R^N $ is the weight vector
and $b_E \in \mathbb R$ is the bias scalar.
FIG.~\ref{FDNN} illustrates the architecture of $\mathcal F_\text{DNN}$.
As this figure makes clear, each layer models the interaction
between the $n$th and $m$th KS orbitals in the vector.

We implement $\mathcal{HK}_\text{DNN}$
described in the main text using a similar feedforward architecture:
\begin{eqnarray}
&& \mathbf h^{(1)}(\mathbf r_j) = \mathbf w_{\rho} \rho(\mathbf r_j) + b_{\rho},
\\
&& \mathbf h^{(\ell+1)}(\mathbf r_j) = \text{ReLU} (
\mathbf W^{(\ell)}_{\text{HK}} \mathbf h^{(\ell)}(\mathbf r_j)
+ \mathbf b^{(\ell)}_{\text{HK}} ),
\\
&& V'_{\mathcal M}(\mathbf r_j) = \mathbf w^{\top}_V \mathbf h^{(L')}(\mathbf r_j) + b_V.
\end{eqnarray}
Thus, $\mathcal{HK}_\text{DNN}$ maps a scalar $\rho$ to another scalar $V$.
In this map, each hidden layer $\mathbf h \in \mathbb R^{N'}$
is an $N'$-dimensional vector, where $N'$ is the number of hidden units
and a hyperparameter of the model.
FIG.~\ref{HKDNN} illustrates the architecture of $\mathcal{HK}_\text{DNN}$.

\begin{figure*}[t]
\begin{center}
\includegraphics[width=17cm, bb=0 0 1920 1080]{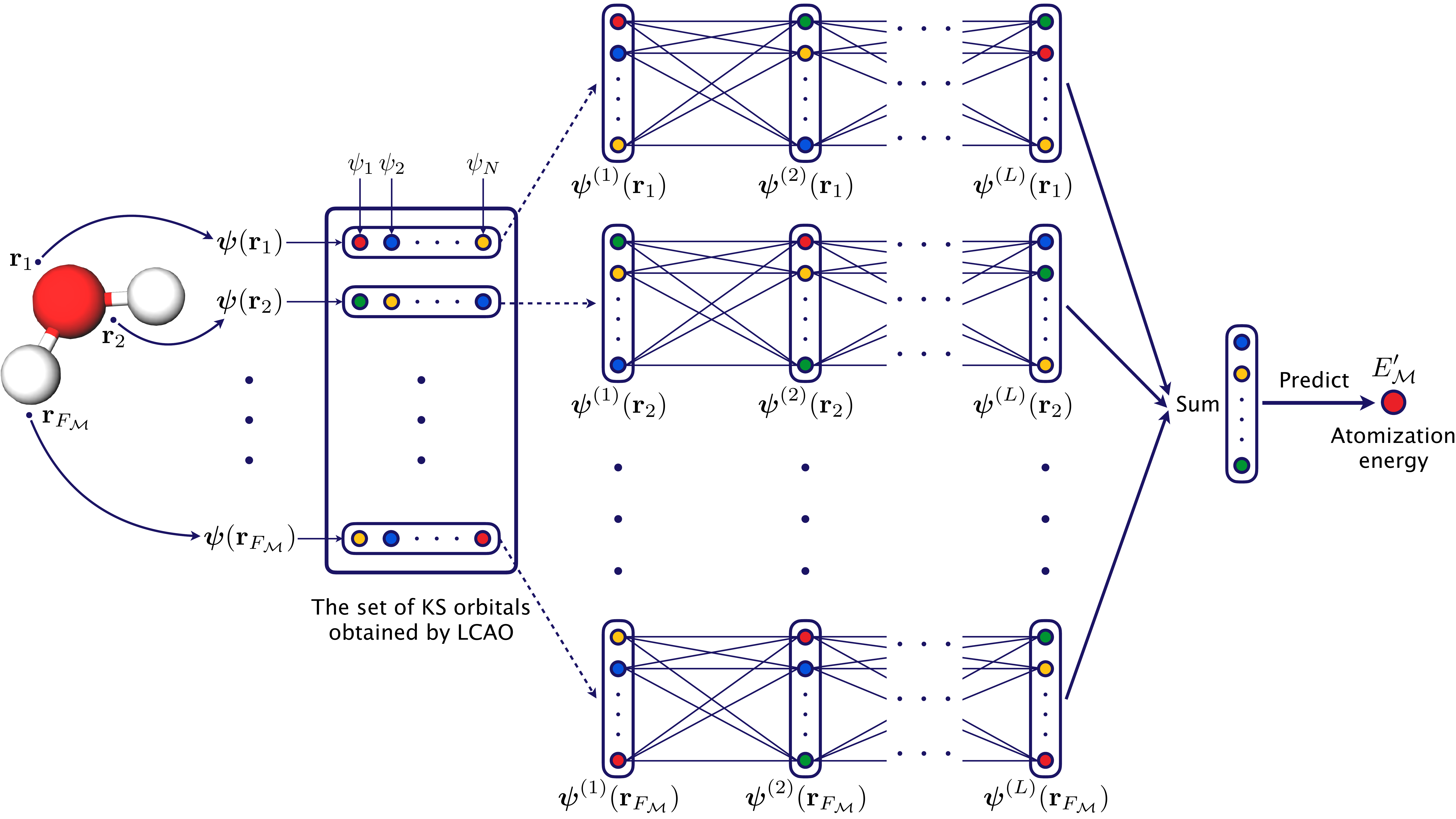}
\caption
{\label{FDNN}
Architecture of the DNN-based energy functional
$\mathcal F_\text{DNN}$ described in the main text.
}
\end{center}
\end{figure*}

\begin{figure*}[t]
\begin{center}
\includegraphics[width=17cm, bb=0 0 1920 1080]{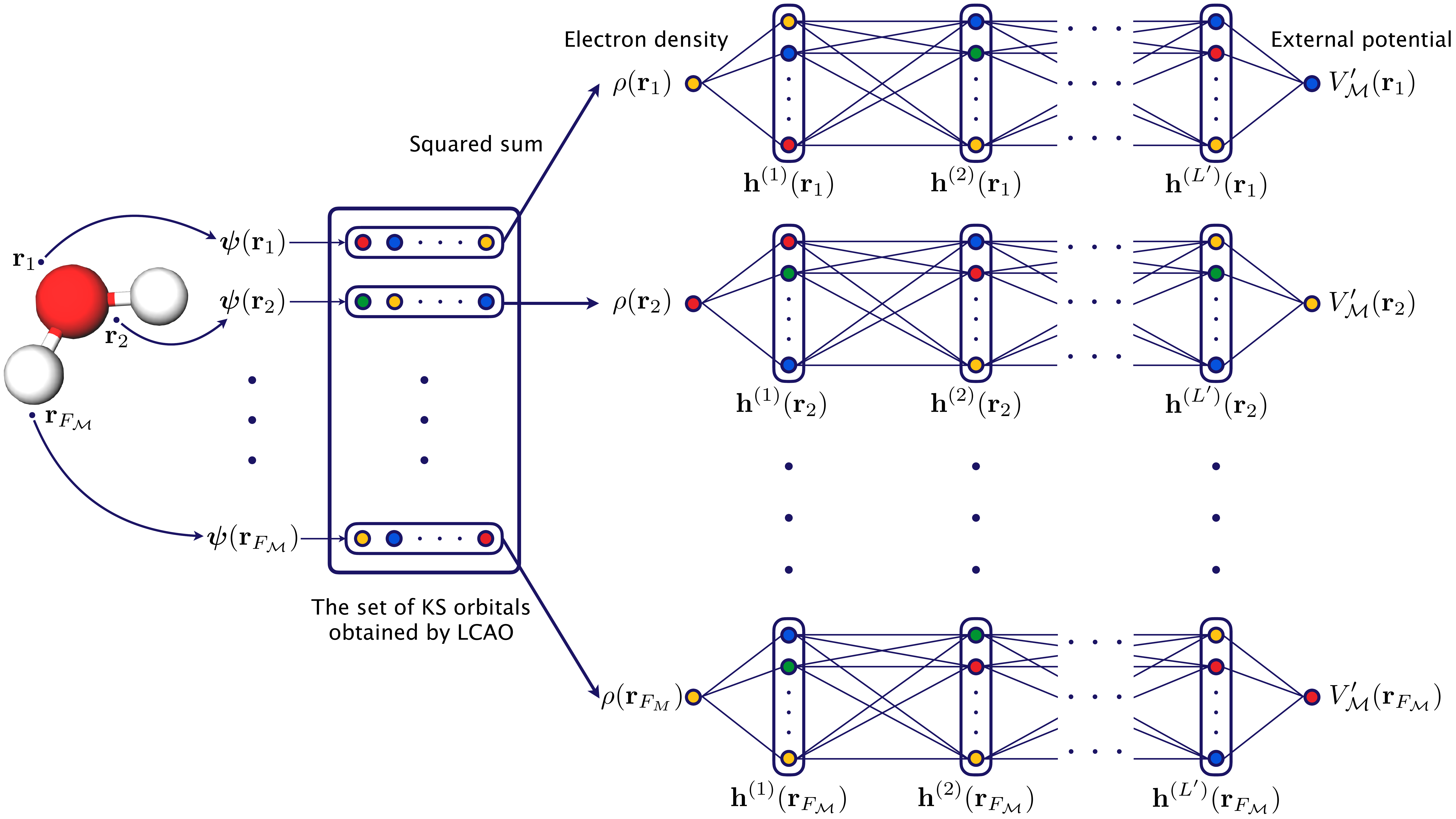}
\caption
{\label{HKDNN}
Architecture of the DNN-based HK map
$\mathcal{HK}_\text{DNN}$ described in the main text.
}
\end{center}
\end{figure*}

\subsection*{Optimization}

Using the backpropagation and an iterative SGD-based learning algorithm,
we minimize the loss function $\mathcal L_E$ in the main text;
in other words, we update the set of learning parameters
$\Theta_E = \{ \{ \zeta_i \}_{i=1}^N, \{ \mathbf c_i \}_{i=1}^N,
\{ \mathbf W^{(\ell)}_E \}_{\ell=1}^L, \{ \mathbf b^{(\ell)}_E \}_{\ell=1}^L,
\mathbf w_E, b_E \}$ as follows:
\begin{eqnarray}
\Theta_E \leftarrow \Theta_E - \alpha \frac{1}{B} \sum_{k=1}^B
\frac {\partial \mathcal L_{E_{\mathcal M_k}}} {\partial \Theta_E},
\end{eqnarray}
where $\mathcal L_{E_{\mathcal M_k}}$ is the atomization energy loss value of
the $k$th molecule $\mathcal M$ in the training dataset,
$\alpha$ is the learning rate, and $B$ is the batch size.
Additionally, we also minimize the loss function
$\mathcal L_V = \int ||V_{\mathcal M}(\mathbf r) - V'_{\mathcal M}(\mathbf r)||^2 d \mathbf r \approx
\sum_{j=1}^{F_{\mathcal M}} ||V_{\mathcal M}(\mathbf r_j) - V'_{\mathcal M}(\mathbf r_j)||^2$,
i.e., we update the set of learning parameters
$\Theta_V = \{ \{ \zeta_i \}_{i=1}^N, \{ \mathbf c_i \}_{i=1}^N,
\mathbf w_{\rho}, b_{\rho}, \{ \mathbf W^{(\ell)}_{\text{HK}} \}_{\ell=1}^{L'},
\\
\{ \mathbf b^{(\ell)}_{\text{HK}} \}_{\ell=1}^{L'},
\mathbf w_V, b_V \}$ as follows:
\begin{eqnarray}
\Theta_V \leftarrow \Theta_V - \alpha \frac{1}{B} \sum_{k=1}^B
\frac {\partial \mathcal L_{V_{\mathcal M_k}}} {\partial \Theta_V},
\end{eqnarray}
where $\mathcal L_{V_{\mathcal M_k}}$ is the external potential loss value of
the $k$th molecule $\mathcal M$ in the training dataset.
In practice, the SGD in this study used the Adam optimizer~\cite{kingma2014adam}.
Note that we update $\Theta_E$ and $\Theta_V$ alternately
and the learning parameters in LCAO, i.e., $\{ \zeta_i \}_{i=1}^N$
and $\{ \mathbf c_i \}_{i=1}^N$, are shared in $\Theta_E$ and $\Theta_V$.

\subsection*{Normalization}

In LCAO, we consider the normalization for the coefficients,
i.e., $\sum_{i=1}^N c_{ni}^2 = 1$.
This can be implemented by updating $\mathbf c'_n$
in the iterative SGD-based learning algorithm as follows:
\begin{eqnarray}
\mathbf c'_n \leftarrow \frac{\mathbf c'_n}{|\mathbf c'_n|},
\end{eqnarray}
where $\mathbf c'_n \in \mathbb R^N$ is not the $n$th row vector (i.e., $\mathbf c_n$)
but the $n$th column vector of the coefficient matrix (see our implementation).
Additionally, the normalization term in GTO is calculated as follows:
\begin{eqnarray}
Z(q_i, \zeta_i)
&=& \int | D_i^{(q_i-1)} e^{-\zeta_i D_i^2} | ^2 dD_i
\\
&=& \sqrt{\frac {(2q_i-3)!! \sqrt{\pi/2}} {2^{2(q_i-1)} \zeta_i^{(2q_i-1)/2}}}.
\end{eqnarray}
Note that because each orbital exponent $\zeta_i$ is a learning parameter of the model,
$Z(q_i, \zeta_i)$ is recalculated every time the model parameters are updated.
Furthermore, we must consider the total electrons: $\int \rho(\mathbf r) d \mathbf r
\approx \sum_{j=1}^{F_{\mathcal M}} \rho(\mathbf r_j)
= \sum_{j=1}^{F_{\mathcal M}} \sum_{n=1}^N | \psi_n(\mathbf r_j) |^2
= N_{\text{elec}}$, where $N_{\text{elec}}$ is the total electrons in $\mathcal M$.
We implemented this by updating $\boldsymbol \psi_n$ in the iterative algorithm as follows:
\begin{eqnarray}
\boldsymbol \psi_n \leftarrow \sqrt{\frac{N_{\text{elec}}}{N}}
\frac{\boldsymbol \psi_n}{|\boldsymbol \psi_n|},
\end{eqnarray}
where $\boldsymbol \psi_n \in \mathbb R^{\mathcal F_{\mathcal M}}$
is not the $j$th row vector (i.e., $\boldsymbol \psi(\mathbf r_j) \in \mathbb R^N$)
but the $n$th column vector of the matrix in FIGs~6 and 7 (see our implementation).

\subsection*{Hyperparameters}

All model/optimization hyperparameters and their values used in this letter
are listed in TABLE~\ref{hyperparameters} (see also our source code).
It is important to note that $N$, which is the dimensionality of coefficient vector
or the number of atomic basis functions in LCAO,
is actually different for each molecule in quantum simulations.
However, an ML model needs to set a common global $N$ for all molecules in a dataset
and the QM9 dataset contains only small organic molecules
made up of less than 30 H, C, N, O, and F atoms.
Considering the maximum size of molecules in the QM9 dataset
and a standard 6-31G basis set, we set $N = 200$.

\begin{table}[t]
\begin{center}
\begin{tabular}{lr}
\hline \hline
Hyperparameter & Value \\
\hline
Sphere radius & 0.75 \AA \\
Grid interval & 0.3 \AA \\
\# of dimensions $N$ & 200 \\
\# of hidden layers in $\mathcal F_\text{DNN}$ & 6 \\
\# of hidden units in $\mathcal{HK}_\text{DNN}$ & 200 \\
\# of hidden layers in $\mathcal{HK}_\text{DNN}$ & 6 \\
Batch size & 4 \\
Learning rate & 1e-4 \\
Decay of learning rate & 0.5 \\
Step size of decay & 300 epochs \\
Iteration & 3000 epochs \\
\hline \hline
\end{tabular}
\end{center}
\caption{
\label{hyperparameters}
The list of all model/optimization hyperparameters used in this study and their values.
}
\end{table}

\subsection*{Limitations and future directions}

Our current QDF framework also comes with some limitations
relative to the DFT calculations and other ML/DNN approaches.

QDF and other ML/DNN approaches using such as
the Coulomb matrix~\cite{rupp2012fast} and SchNet~\cite{schutt2018schnet}
require a 3D (e.g., the DFT-relaxed) structure of molecule in the first place;
in other words, these do not involve the molecular structure optimization process
and cannot be a complete replacement of the DFT calculations.
We believe the development of ML/DNN models including
the molecular structure optimization process is interesting
but very challenging in terms of learning the model
and optimizing the structure; we leave it for future work.

To avoid the use of the 3D structure
required in the Coulomb matrix, SchNet, and our QDF,
we can use the molecular descriptors based on chemical composition data.
Indeed, Faber et al., 2017~\cite{faber2017prediction} conducted a comprehensive
evaluation of various ML models with various descriptors on the QM9 dataset.
In this evaluation, the prediction performance for the atomization energy
of the kernel ridge regression (KRR) as the ML model
and the molecular fingerprints as the descriptor was very poor;
the error was 4.25 eV = 98.0 kcal/mol.
Faber et al. also reported that the performance of the GNN~\cite{kearnes2016molecular}
using the molecular graph was significantly better than that of KRR.
However, the molecular graph is based on the adjacency (i.e., binary) matrix
describing the chemical bonds (e.g., \ce{C=O} and \ce{N-H}); this implies that
GNNs consider the atomic wave/basis functions as the binary (i.e., 0 or 1) values.
Additionally, the deep nonlinearity in GNNs destroys the linearity in LCAO.
Furthermore, as we have described in the main text, these characteristics
degrade the extrapolation performance compared to our QDF.
From these observations, we believe that
the 3D structure, not the fingerprint and the graph, of molecules
is required to improve prediction and extrapolation performance.

Thus, while it is better to achieve high prediction and extrapolation performance
using only the fingerprint and the graph in terms of cost and simplicity,
current ML models struggle to achieve this without the 3D structure.
However, we believe that our QDF can be adapted to the molecular graph
based on the atomic distances focused on the chemical bonds obtained by
SMILES and a software such as RDKit (\url{https://www.rdkit.org/}).
Using the atomic distances or chemical bond lengths,
our QDF can learn the atomization energy; this will be a ``compromise''
between the current QDF with the DFT-relaxed 3D structure
and the GNN with the molecular graph, which can be called a graph-based QDF.
We believe that if the graph-based QDF achieves reasonable performance
compared to the current QDF, our QDF framework can be used as a practical
high-throughput prediction method without the 3D structure.

While our implemented QDF model fulfills the normalization condition in
the atomic and molecular orbitals (precisely, the basis functions and the KS orbitals),
the model does not satisfy the orthogonality of the orbitals.
We believe that it is difficult for an ML model to consider
the orthogonality in terms of learning cost, e.g., we would additionally
need to learn the orthogonality for all orbital pairs.
Fortunately, however, we obtained qualitatively valid electron density
from the current orbitals, so this study did not consider the orthogonality.

In QDF, the set of orbital exponents $\{ \zeta_i \}_{i=1}^N$
and coefficient vectors $\{ \mathbf c_i \}_{i=1}^N$ of LCAO
are the common global learning parameters for all molecules;
on the other hand, these actually differ for each real molecule
(i.e., characterized with the atomic environment) in quantum simulations.
This is considered in GNN variants (e.g., the interaction blocks in DTNN and SchNet)
and ignoring this characteristic may be a limitation of the current QDF.
However, we also believe that such global parameters
prevent the model from being too flexible for each molecule.
Considering this, our implementation would be reasonable in terms of
reducing the model parameters and complexity, leading to robust extrapolation.

The selection of the basis set is critical for the prediction accuracy of QDF,
which is the same as for the computational accuracy of DFT.
This study used the 6-31G basis set considering the level of theory of the QM9 dataset;
of course, QDF can use other basis sets, such as the 6-311G, 6-31+G(d,p),
and STO not GTO, for improving the performances of
atomization energy prediction/extrapolation and electron density generation.
Additionally, although this study focused on the isolated molecule,
for modeling the crystal structure QDF must use the plane wave basis set.
Recently, the crystal graph convolutional neural network (GCNN)
has been proposed~\cite{xie2018crystal} and successfully learned and predicted
the formation energy, band gap, and other properties of crystals of over 60,000 samples
in the Materials Project database~\cite{jain2013commentary}.
The crystal GCNN and its variants, however, do not consider
the plane waves and LCAO (i.e., tight-binding approximation);
we believe that QDF can be extended to crystals and will replace the GCNN.
In the crystal modeling, the selection of the plane wave basis set
is also critical for the prediction accuracy of QDF.

\begin{figure*}[t]
\begin{center}
\includegraphics[width=17cm, bb=0 0 1920 742]{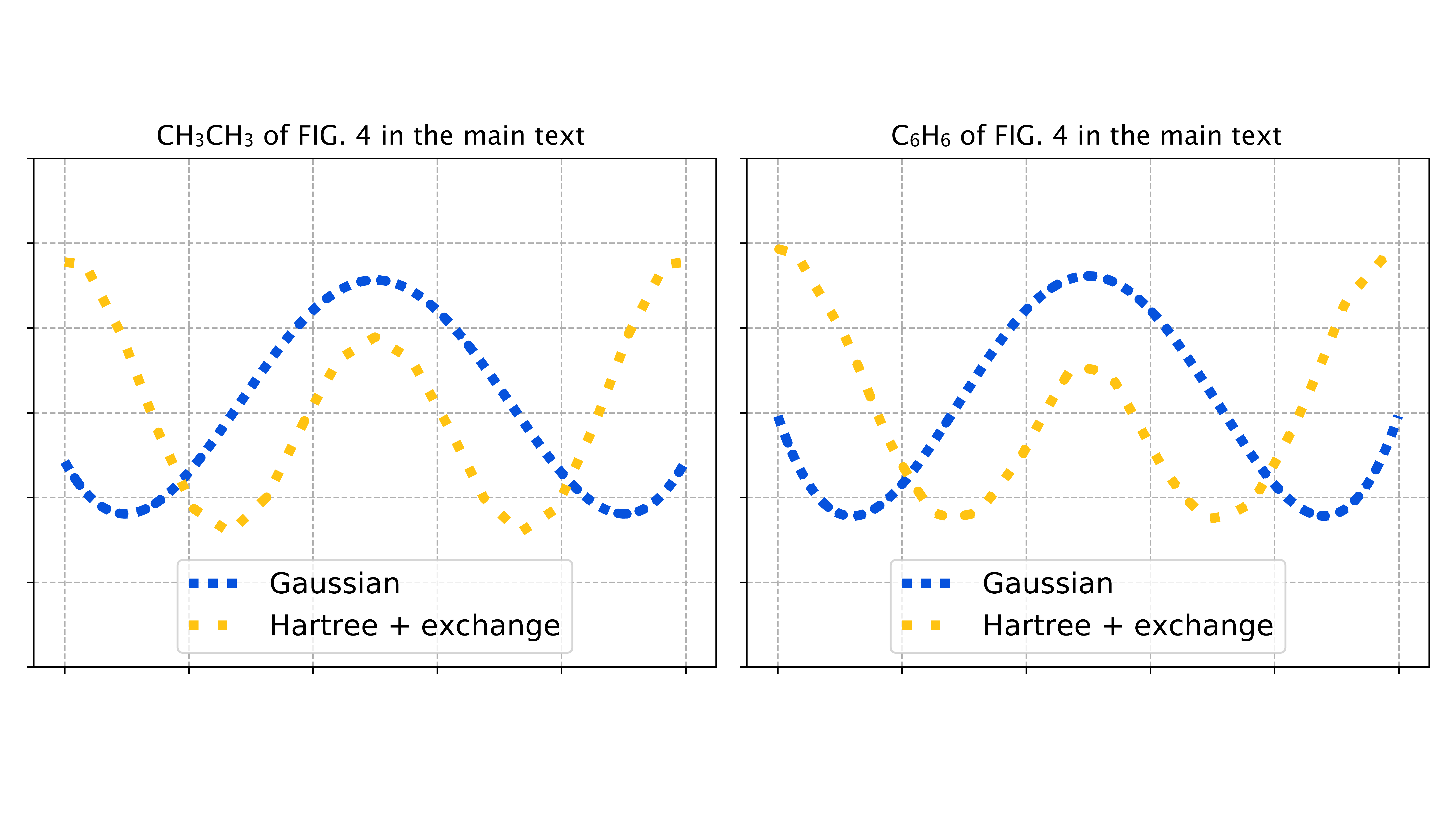}
\caption
{\label{potential}
Comparison of the potentials.
}
\end{center}
\end{figure*}

The nonlinear components in the current implementation
are most straightforward; in other words, the model used in this study
is the simplest special case of the QDF framework.
As shown in FIGs~\ref{FDNN} and \ref{HKDNN},
we implemented $\mathcal F_\text{DNN}$ and $\mathcal{HK}_\text{DNN}$
by a feedforward architecture; however, these can be further extended.
Indeed, both DNNs do not use any additional architectures and techniques,
e.g., residual networks, drop out, and batch normalization,
and these will be effective for QDF.
Additionally, the current HK map is $V = \mathcal{HK}_{\text{DNN}}(\rho)$,
that is, this is a local density approximation (LDA) fashion
and this study used it as a first step.
We emphasize that even using this LDA-like HK map, we achieved high accuracy
in terms of predicting/extrapolating the atomization energy
and generating the electron density. Of course, this can be improved.
For example, instead of $V = \mathcal{HK}_{\text{DNN}}(\rho)$,
we can consider a generalized gradient approximation, i.e., the GGA-like HK map
$V = \mathcal{HK}_{\text{DNN}}(\rho, \nabla \rho)$.
Furthermore, as we have described in the main text
(see also comparison of the potentials in FIG.~\ref{potential}),
the Gaussian external potential should also be improved
and this can be addressed relatively easily because
we only consider and set another external potential and learn it as a new target.

QDF requires much more training time and memory than GNN,
because of the 6-31G basis set and all points in the fine grid fields of molecules.
Indeed, GNN can use a large batch size (e.g., $B = 64$ and $B = 128$);
however, QDF can use $B = 4$ at most and this did not allow us to
train the model on all 130,000 molecules of the QM9 dataset.
We believe that training a QDF model on a larger number of data samples
will require the efficient use of dozens of GPUs.

Although QDF has the above limitations
(e.g., as a replacement model of the DFT calculations),
the important point is that QDF has many extensions and applications;
in particular, we believe that QDF will provide transfer learning applications
to solve practical problems in materials informatics.
For example, the pre-trained QDF model with the atomization energy
of the large-scale QM9 dataset can be used for transfer learning
and predicting of other properties (e.g., the HOMO, LUMO, and their gap) universally.
The reason is that the model can encode information about
the fundamental quantum characteristics (i.e., $\psi$ and $\rho$) of molecules
by learning the HK map constraint, which is shown by
the electron density comparison with the DFT calculation
and the extrapolation evaluation in predicting the atomization energy of large molecules.
We now plan to transfer the pre-trained QDF model to other datasets and properties
such as the HOMO--LUMO gap of much larger molecules, i.e., polymers~\cite{huan2016polymer},
extend the current implementation to crystal structure data
in the Materials Project database~\cite{jain2013commentary},
and evaluate the extrapolation in predicting the polymer and crystal properties.

\subsection*{Data and code availability}

Our QDF implementation in PyTorch, including the preprocessed QM9 dataset,
is available at \url{https://github.com/masashitsubaki}.
Model extensions can be created by forking this source code.

\end{document}